\documentclass[11pt]{article}
\usepackage{graphicx}
\usepackage{amsfonts}
\usepackage{amssymb}
\usepackage[centertags]{amsmath}
\usepackage{amsfonts,bm}
\usepackage{newlfont}
\usepackage{graphicx}
\usepackage{eurosym}
\usepackage{xcolor}
\usepackage{epstopdf}
\usepackage{caption}
\usepackage{subcaption}
\newcommand{\Proof}{{\em Proof: }}

\newtheorem{Theorem}{Theorem}[section]
\newtheorem{Proposition}[Theorem]{Proposition}

\newtheorem{Remark}[Theorem]{Remark}
\newtheorem{Lemma}[Theorem]{Lemma}
\newcommand{\E}{\mathbb{E}}
\newcommand{\bP}{\mathbb{P}}
\newcommand{\Q}{\mathbb{Q}}

\newcommand{\R}{\mathbb{R}}

\newcommand{\e}{\mathrm{e}}
\newcommand{\lam}{\lambda}
\newcommand{\eps}{\epsilon}
\newcommand{\ds}{\displaystyle}
\newcommand{\cA}{\cal A}

\newcommand{\cF}{\cal F}
\newcommand{\cG}{\cal G}
\newcommand{\cH}{\cal H}

\newcommand{\cN}{\cal N}

\newcommand{\brho}{\bm{\rho}}

\begin{document}

\title{Approximate XVA for European claims}

\author{F. Antonelli\footnote{ University of L'Aquila,
          \texttt{fabio.antonelli@univaq.it}}, A. Ramponi\footnote{Dept. Economics and Finance, University of Roma - Tor Vergata, \texttt{alessandro.ramponi@uniroma2.it}}, S. Scarlatti\footnote{Dept. Enterprise Engineering, University of Roma - Tor Vergata, \texttt{sergio.scarlatti@uniroma2.it}}}


\maketitle

\begin{abstract}
We consider the problem of computing  the Value Adjustment of  European contingent claims when default of either party is considered, possibly including also   funding and collateralization requirements. 

As shown in  Brigo et al. (\cite{BLPS}, \cite{BFP}), this leads to a more articulate variety of Value Adjustments  ({XVA}) that introduce some nonlinear features. When exploiting a reduced-form approach for the default times, the adjusted price can be characterized as the solution to a possibly nonlinear Backward Stochastic Differential Equation (BSDE). The expectation representing the solution of the BSDE is usually quite hard to compute even in a Markovian setting, and one might resort either to the discretization of the Partial Differential Equation characterizing it or to Monte Carlo Simulations. Both choices are computationally very expensive and in this paper we suggest an approximation method based on an appropriate change of numeraire and on a Taylor's polynomial expansion when intensities are represented by means of affine processes correlated with the asset's price. The numerical discussion at the end of this work shows that, at least in the case of the CIR intensity model, even the simple first-order approximation has a remarkable computational efficiency.

\medskip

\end{abstract}

\section{Introduction}
\label{intro}

Many financial institutions trade contracts in over-the-counter (OTC) markets, their counterparties  being other financial institutions or corporate clients. However, many of those contracts are subject, to some extent, to counterparty risk, or in other words, they are subject to some default event concerning the solvency of either one of the parties, that might take place during the lifetime of the contract. These are called defaultable.  Initially, the evaluation regarded European options, named vulnerable, when the seller's default was the only risk and two approaches emerged over the years: the structural approach and the reduced form approach. 

Historically, the structural approach came first introduced by  Johnson and Stulz  in \cite{JS87} when they
 considered  the option as the sole liability of the counterparty. In the same framework, in \cite{Kl96}  Klein discussed  more general liability structures, in \cite{KIn99} he included interest rate risk, and in   \cite{KIn01}  he considered  a (stochastic) default barrier depending on the value of the option.  More recently, \cite{TW14} extended this approach to jump-diffusion models,  \cite{Kao16} considered multiple correlations, \cite{CL03} treated it by using copulas.
 
Then researchers developed the alternative  reduced-form approach.  For a comprehensive presentation of the topic, we refer the reader to  \cite{Lando}. In  \cite{DS99},  and the references therein, one can find a general overview of the approach for defaultable bonds.  Later, the approach's mathematical framework was carefully formalized in \cite{BR} and \cite{BJR}, and recently  \cite{CPV14} and \cite{Fard15} extended it to defaultable claims  in Levy  market models.

In the last decade,  after the financial crisis of 2008-09, the interest in {Counterparty Credit Risk}  increased remarkably, 
and attention focused on building a general framework  to define and evaluate the premium to compensate the risk connected to defaultable products (in particular of Interest Rate Swaps). This premium took the name of  Credit Value Adjustment ({CVA}) in the seminal paper by Zhu and Pykhtin \cite{ZP07}, and it defines  the appropriate reduction of  the default-free value of a portfolio, to compensate for the default risk. This discount became the crucial quantity to take into account when trading derivatives in OTC markets, spurring much research in the field: see, for instance,  \cite{BCB}, \cite{BMP13}, \cite{Gr12}. 

Over the years, other value adjustments were introduced in the contract's evaluation, leading to the acronym (X)VA. Here,  X stands for D= debt, L= liquidity, F=funding, to include also the risks due to the default of both parties, funding investment strategies, lack of liquidity.  We refer the reader to \cite{Gree16} for a comprehensive exposition on the matter. In \cite{GSZ17}, one might find an updated overview of the recent research directions under investigation. We point out that there the characterization of the adjusted value as the solution of a BSDE is very well explained. In a Markovian setting, the connection between bilateral CVA and  Partial Differential Equations (PDEs) is also thoroughly investigated in \cite{BK11} and further developed in \cite{BK17}.is 

In this work, we  treat a European claim, whose price is influenced by the default probabilities of either party as well by liquidity, financing, and collateralization risks when exploiting the intensity approach for the default times of both parties. 

In a remarkable series of papers, (\cite{BLPS}, \cite{BFP}, \cite{BFP1}),  Brigo et al. describe in detail how introducing all the value adjustments implies the loss of an explicit expression for the adjusted value.  Indeed the BSDE characterizing the contract's value is generally nonlinear and hence hardly solvable. It depends on the asset's price and many other, possibly correlated, factors such as default intensities, interest rate, stochastic volatility, so that even in a Markovian setting, the expectation representing the solution of the associated PDE becomes extremely difficult to evaluate. Hence to provide a numerical approximation, one may resort only to the discretization of the PDE characterizing the solution of the BSDE (see \cite{KL16}) or to Monte Carlo simulations (as in \cite{BFP}). Either approach, on average computational resources,  results to be computationally very expensive.

We are interested in devising an approximation procedure simple and computationally efficient even in the presence of many stochastic factors, provided we make some modeling choices.  Indeed, 
we suggest  to view the evaluation expectation as a smooth function of the correlation parameters and to approximate it by its Taylor polynomial expansion around the zero vector (the independent  case), in the hope that the first or second-order are enough to provide an accurate approximation.
We apply our method to estimate the price contribution that comes from considering stochastic default intensities correlated with the underlying's price. We remark, though, that we can straightforward extend the same technique to include further stochastic factors.

To evaluate Taylor polynomial's coefficients, we follow a two-step procedure to exploit, whenever possible, explicit formulae from option and bond's pricing theory. First, we condition the underlying's price with respect to the stochastic factors, retrieving a conditional Black  \& Scholes formula.  Then, assuming the intensities to be described by affine models, we represent the single terms of the expansion using a change of Numeraire technique (similar to the one in \cite{BR18}) to disentangle the correlation among the asset's price and the default intensities. The affinity of the processes makes it possible to use a ``bond-like" expression for the default component.

To carry out the calculations in detail and to perform the numerical analysis of the method, we represent the intensities by two Cox Ingersoll Ross (CIR) processes. The final section shows the method's efficiency using Monte Carlo simulations as a benchmark.

A strong point of this approach is that it provides a relatively simple method that one can use with many correlated processes. Correlation often destroys any affine property the dynamical system might have, making the Riccati equations/Fourier transform framework inapplicable, and one can resort only to Monte Carlo or PDE's approximations.
The latter are both computationally expensive in several dimensions, hence the construction of an alternative with a remarkable gain in computational time, without loss in accuracy, becomes very important.
 
Our method becomes particularly convenient when the correlation structure (as Monte Carlo simulations point out for the CIR model) seems to follow a linear pattern.  In this case, a first-order Taylor's polynomial is enough to produce an accurate approximation, providing a rather handy evaluation formula. 
We finally remark that the conditioning and change of numeraire techniques allow us to keep the coefficients' approximations to a minimum. The expansion's zeroth term corresponds to the independent case,  and we need to have a semi-explicit formula to evaluate it.  This fact forced us to restrict our model choices.

The paper is structured as follows.  In the next section, we describe the general problem leading to the BSDE characterization under the reduced-form approach. We specify the model and the two-step evaluation procedure to compute Taylor's approximation in Section 3, while in Section 4, we specialize the calculations when the default intensities are CIR processes. Section 5 concerns the numerical analysis of our results.

\section{XVA Evaluation of European claims under the intensity approach}

We consider  a finite time interval  $[0,T]$ and a complete probability space $(\Omega, \cF, \bP)$, endowed with a filtration  $\{\cF_t\}_{t\in [0,T]}$, augmented with the
$\bP-$null sets and made right continuous. We assume that all  processes have a c\'adl\'ag version.

The market is described by the interest rate process $r_t$ determining the money market account
and by an adapted process  $X_t$  representing an asset log-price (we will specify its dynamics later), which may also depend on additional stochastic factors.
We assume

\begin{itemize}
\item that the filtration $\{\cF_t\}_{t\in [0,T]}$ is rich enough (and possibly more) to support all the stochastic processes that describe the market;

\item to be  in absence of arbitrage;

\item that the given probability  $\bP$ is a risk-neutral measure, already selected by some criterion.

\end{itemize}

In this market model (as in  \cite{BFP}) we consider two parties ($I=$ investor, $C=$ counterparty) exchanging some European claim with default-free payoff  $f(X_T)$,  where $f$ is a function (not necessarily nonnegative) as regular as needed. We take for granted that the market processes fulfill the necessary integrability hypotheses to guarantee a good definition of all the expectations we are going to write.

Both parties might default, due to some critical credit state, with respective random times $\tau^1$ (Counterparty) and $\tau^2$ (Investor), which are not stopping times with respect to the filtration $\cF_t$. In this context we define
the filtration $\cG_t=\cF_t\lor\cH_t^1\lor \cH_t^2$, where  $\cH_t^i =\sigma( \mathbf 1_{\{\tau^i\le s\}}, s \leq t)$, $i=1,2$, which is the smallest filtration extension that makes both random variables stopping times. Moreover, we assume  there exists a unique extension of the risk-neutral probability to $\cG_t$, that we keep denoting by $\bP$.

In general, the following fundamental assumption, known as the  H-hypothesis (see e.g. \cite{GJLR10} and \cite{G14} and the references therein), ensures price coherence:

\medskip

\noindent
(H)\qquad \qquad \qquad  Every $\cF_t-$martingale remains a $\cG_t-$ martingale.\hfill

\medskip
\noindent 
By Lemma 7.3.5.1   in \cite{JYC09}, (H) is automatically satisfied, under square integrability of the payoff, by the default-free price of any European contingent claim, whence we may affirm that
$$
\begin{aligned}
\e^{\int_0^tr_u du}  \e^{X_t}&= \E(\e^{\int_0^T r_u du}   \e^{X_T}|\cF_t)=\E(\e^{\int_0^T r_u du}  \e^{X_T}|\cG_t)\\
\e^{\int_0^t r_u du} c(t,T) &:=\E(\e^{\int_0^T r_u du} f(X_T)|\cF_t)=\E(\e^{\int_0^T r_u du} f(X_T)|\cG_t)
\end{aligned}
$$
remain  $\cG_t-$ martingales under  $\bP$, for all $t\in [0,T]$.

In what follows, to stress the significance of the term ``adjustment", we will point the corrections out step by step, with their signs determined by the fact that we are taking the investor?s viewpoint. 

We start assuming full knowledge that is we are in the  $\cG_t-$ filtration.
The contract makes sense only if the default of either party has not occurred yet at the evaluation time $t$. Denoting by $\tau=\min(\tau^1, \tau^2)$, this fact is represented by the indicator function $\mathbf 1_{\{\tau>t\}} $ to be placed in front of the price.

Either party may  default, so a bilateral adjustment is needed. For the moment  we assume nothing is recovered at  default. Denoting by CVA$^0(t,T)$ the Credit Value Adjustment due to the counterparty's default, this quantity has to act as a discount to the default-free price to balance the investor's risk assumption. On the other hand, the Debt Value Adjustment due to the investor's default,   DVA$^0(t,T)$, has to act as an accrual of the default-free price as it compensates the counterparty's risk assumption. So,  for the $\cG_t-$adapted  adjusted value of the European claim  $c^\cG(t,T)$, we may write
\begin{equation}
\label{value0}
\mathbf 1_{\{\tau>t\}} c^\cG(t,T)=\mathbf 1_{\{\tau>t\}} \Big [c(t,T)- \text{CVA}^0(t,T)+ \text{DVA}^0(t,T)\Big ],
\end{equation}
where CVA$^0(t,T)$ and DVA$^0(t,T)\ge 0$.

Now, let us admit the defaulting party might partially compensate for the loss due to his/her 
default.  In this case, we have to include other two nonnegative
 terms, CVA$^{rec}(t,T)$ and  DVA$^{rec}(t,T)$ (respectively for the counterparty and the investor), and we can rewrite the above as 
\begin{equation*}
\label{value1}
\mathbf 1_{\{\tau>t\}}c^\cG(t,T)=\mathbf 1_{\{\tau>t\}} \Big [c(t,T)- \text{CVA}^0(t,T)+\text{DVA}^0(t,T) + \text{CVA}^{rec}(t,T)-\text{DVA}^{rec}(t,T)\Big ].
\end{equation*}
Moreover,  as explained in \cite{BFP1}, the two parties might  be asked to collateralize their participation to the contract, they might need  to borrow money to finance this participation and/or the risky asset(s) from a  repo market to rea,lize their hedging strategies. All this leads to funding and liquidity risks that, again, have to be included for the correct contract's evaluation. Thus, we should write
\begin{equation}
\label{value2}
\begin{aligned}
\mathbf 1_{\{\tau>t\}}c^\cG(t,T)=&\mathbf 1_{\{\tau>t\}}\Big [c(t,T)-\text{CVA}^0(t,T)+\text{DVA}^0(t,T)\\
 &+ \text{CVA}^{rec}(t,T)-\text{DVA}^{rec}(t,T)
+  \text{FVA}(t,T)+ \text{LVA}(t,T)\Big ],
\end{aligned}
\end{equation}
with FVA$(t,T)$, LVA$(t,T)\in \mathbb R$.The first represents the Funding Value Adjustment, the second the Liquidity Value Adjustment, and they are both determined by  strategy financing and collateralization. 

It is then necessary to model these terms to get to a manageable formula. The range of possible choices of mechanisms to include  in the formation of prices is quite broad, and we refer the reader  again to   \cite{BLPS}, \cite{BFP}  and \cite{BFP1} for a
detailed discussion. Of course, there is an interplay among the different cash flows. For instance, collateralization changes the parties? exposures, the amount of cash borrowed at rate r increases its value at a rate $r_s$.

Here we use  the following set of assumptions.

\begin{enumerate}

\item The claim pays no dividends.

\item  The adjustment
processes all depend on  a close-out value, $\eps_t$, determined by a contractual agreement. It is natural to consider  it $\cF_t-$adapted since it is established on the basis of the information before default. Usually, it is taken as the default-free price or as the price of the defaultable claim itself.

\item We denote the collateralization process by $C_s$ and   it is  a, possibly time-varying, percentage of the close-out value
\begin{equation}
\label{coll}
C_s = \begin{cases}
&\alpha_s \eps^+_s , \quad\text{when due by the counterparty}\\
&\alpha_s \eps^-_s , \quad\text{when due by the investor}
\end{cases}
\qquad 0<\alpha_s<1, \quad \forall s\in [0,T].
\end{equation}
Thus the net exposure is $(\eps_s-C_s)^+=(1-\alpha_s) \eps_s^+$ for the investor and $(\eps_s-C_s)^-=(1-\alpha_s) \eps_s^-$ for the counterparty.

Moreover, we assume that collateralizing happens at rate $r^c_s$.

\item  We denote by $R_1(s)$ the recovery percentage  of the close-out value in case of  counterparty 's default   and by $R_2(s)$, when investor's default occurs. Mirror-like
we define the Loss Given Default as $L_i(s)= (1- R_i(s))$, $i=1,2$.

\item  To build investing strategies, the parties may invest in the riskless asset at a rate $r^\phi$ and the risky asset(s) at a rate $h_t$, the latter happening in a parallel repo market. We denote by  $\phi_u$ the quantity of riskless asset the contract globally requires (either positive or negative) and by $H_t$ the value of the portion of the risky asset(s)  (either positive or negative) traded on the repo market.

Since at the same time the investor's purchase generates wealth at  a rate $r_s$, and as well the borrow/sale of the risky asset generates wealth at a rate $r^\phi$, also this aspect will have to be taken into account.

\end{enumerate}

\begin{table}[t]
\centering
\begin{tabular}{cccc}
\hline
Symbol & Definition & Symbol & Definition \\ \hline
$r_t$ & Risk-free rate & $\tau_1$ & Default time Counterparty  \\ 
$r_t^{\phi}$ & Funding rate &$\tau_2$ & Default time Investor \\
$r_t^c$ & Collateral rate & $\epsilon_t$ & Close-out  value\\
$h_t$ & Hedging rate & $\lambda_t^{i}$& Default intensities \\
$\alpha_t$ & collateralization level & $f(\cdot)$ & Option payoff \\
$R_i(t)$ & Recovery rates $i=1,2$ & $\bar v_t$ & $\int_t^T v_s ds$\\
$\tilde r_t$ & $r^{\phi}_t-h_t$ & $\hat r_t$ & $r^{\phi}_t - r^c_t$ \\ \hline
\end{tabular}
\caption{Summary of notations.}
\label{Tab0}
\end{table}

As we said, the recovery and the collateral agreements are usually a fraction of the close-out value,  and therefore they should be $\cF_t-$adapted. On the contrary, the funding  and hedging processes  $(\phi,H)$ might incorporate the contribution of the default events, and therefore they could be a priori $\cG_t-$adapted.

Finally, the price should be given by the three components 
\begin{equation}
\label{tot}
c^\cG(t,T) = \phi_t+ H_t+ C_t.
\end{equation}
Following the crystal clear exposition  in \cite{BFP} (but also in  \cite{BLPS} and \cite{BFP1} ),
keeping in mind hypothesis (H) and \eqref{tot}, one can obtain the following BSDE in the $\cG-$filtration
\begin{equation}
\hspace{-0.1cm}
\begin{aligned}
&\mathbf 1_{\{\tau>t\}}c^\cG(t,T)=\mathbf 1_{\{\tau>t\}} \Bigg \{\E\left [\e^{-\int_t^T r_udu} f(X_T) \mathbf 1_{\{\tau>  T\}}\big |\cG_t\right] \\
&+\E\left [\e^{-\int_t^\tau \!\!r_udu}\mathbf 1_{\{\tau \le T\}} \Big ( \eps_\tau
\!- \! (1-\alpha_\tau) \big [ L_1(\tau)\eps_\tau^+\mathbf 1_{\{ \tau^1 =\tau\}}\! -\! L_2(\tau) \eps_\tau^-\mathbf 1_{\{ \tau^2 =\tau\}}\big  ]\Big )\Big|\cG_t\right]\\
&+ \left [ \int_t^{\tau\wedge T}\e^{-\int_t^s r_udu} \Big \{ [r_s-r^\phi_s]c^\cG(s,T) ds + [r^\phi_s-r^c_s]C_s+[h_s-r_s] H_s\Big \}ds \Big |\cG_t\right]\Bigg \}.
\end{aligned}
\end{equation}
The random variables  $\tau^i, i=1,2 $,  are not $\cF_t-$stopping times, hence  the traders can observe   only  whether the default events  happened or not, conditioned to the available information.
Thus, any risk-neutral evaluation that would naturally take place in the $\cG-$filtration, needs translating in terms of $\{\cF_t\}$. For that, we have  the following  well known Key Lemma, to be found in  \cite{BJR} or \cite{BCB}, just to quote some references.

\begin{Lemma}\label{key} Given a $\cG_t-$stopping time $\tau$, for any integrable $\cG_T-$measurable r.v. $Y$, the following equality holds

\begin{equation}
\label{key}
\E\Big[\mathbf 1_{\{\tau>
t\}}Y|\cG_t\Big]=\mathbf 1_{\{\tau>t\}}\frac{\E\Big[\mathbf 1_{\{\tau>
t\}}Y|\cF_t\Big]}{\bP(\tau>t|\cF_t)}.
\end{equation}
\end{Lemma}

This Lemma calls for the conditional distributions of the default times that we are going to treat within the (Cox) reduced-form framework. We denote the conditional distribution of the random times as
\begin{eqnarray}
F^i_t=\bP(\tau^i\leq t|\cF_t), \qquad  i=1,2 \qquad\forall \, t\ge 0,
\end{eqnarray}
and  we assume that they both verify $F^i_t<1$. Hence we can define  the corresponding $\cF$- hazard processes of  the $\tau^i $'s as
\begin{equation}
\label{risk}
\Gamma^i_t:=-\ln(1-F^i_t)\quad\Rightarrow \quad F^i_t= 1 - \mathrm e^{-\Gamma^i_t}\quad \forall \, t>0, \qquad  \Gamma_0=0,
\end{equation}
which we assume to be differentiable,  defining the  so-called $\cF_t-$adapted intensity processes $\lam^i$ by
$$
\Gamma^i_t = \int_0^t \lambda^i_u du \quad \Rightarrow \quad F^i_t= 1 - \e^{-\int_0^t \lam^i_udu}.
$$
As in the classical framework of \cite{DH}, we assume conditional independence for the default times, i.e. for any $t > 0$ and
$t_1, t_2\in [0,t]$
$$
\bP(\tau^1 >t_1, \tau^2 > t_2|\cF_t)=  \bP(\tau^1 >t_1|\cF_t)\bP( \tau^2 > t_2|\cF_t),
$$ 
so that we may conclude that $\lambda_t:=\lambda^1_t+\lambda^2_t$ is the intensity process of $\tau=\inf\{\tau^1,\tau^2\}$.

\begin{Remark}
It is worth noting that the independence assumption certainly simplifies computations, but it does not take into consideration default contagion effects. Within the intensity framework, more realistic models allowing default dependence were recently proposed  (see \cite{BCC19}, \cite{BC19} and the references therein),  and we remark that we could extend our method to the correlated case, provided we introduce an additional parameter.
\end{Remark}

Exploiting  the key Lemma and the intensity processes as in \cite{ARS2}, the above equation gets projected on the smaller filtration, obtaining
\begin{equation}
\label{values}
\begin{aligned}
&\mathbf 1_{\{\tau>t\}}c^\cG(t,T) =\mathbf 1_{\{\tau>t\}}\E\Big [
\e^{-\int_t^T( r_u+ \lambda_u)du} f(X_T) \\
&+\int_t^T\e^{-\int_t^s (r_u+\lambda_u)du} \big [\lambda_s \eps_s
- (1-\alpha_s)\big (\lambda^1_s L_1(s)\eps_s^+ - \lambda^2_sL_2(s)\eps_s^- \big ) \big ]ds\\
&+ \int_t^T\e^{-\int_t^s (r_u+\lambda_u)du}  \big [\big(r_s-r^\phi_s\big)c^\cG(s,T) +\big(r^\phi_s-r^c_s\big )\alpha_s \eps_s + (h_s-r_s) H_s\big]ds\Big  |\cF_t\Big].
\end{aligned}
\end{equation}
Applying the Key Lemma  and Lemma 2 in   \cite{BFP}  (extension of the key lemma) to \eqref{values}, we may conclude that there exists an $\cF_t-$adapted  adjusted price of the European claim,  $c^a(t,T)$ and an adapted hedging strategy (the part hedging the default-free risks) $\tilde H$ such that
$$
c^a(t,T)\mathbf 1_{\{\tau>t\}}=c^\cG(t,T)\mathbf 1_{\{\tau>t\}},\quad \tilde H_t\mathbf 1_{\{\tau>t\}}=H_t\mathbf 1_{\{\tau>t\}},
$$
and we may conclude that  on $\{\tau>t\}$ 
\begin{equation}
\label{BSDE1}
\begin{aligned}
&\mathbf 1_{\{\tau>t\}}c^a(t,T) =\mathbf 1_{\{\tau>t\}}\E\Big [
\e^{-\int_t^T( r_u+ \lambda_u)du} f(X_T) \\
&+\int_t^T\e^{-\int_t^s (r_u+\lambda_u)du}  \big [\lambda_s \eps_s
- (1-\alpha_s)\big (\lambda^1_s L_1(s)\eps_s^+ - \lambda^2_sL_2(s)\eps_s^- \big ) \big ]ds\\
&+ \int_t^T\e^{-\int_t^s (r_u+\lambda_u)du} \big  [\big ( r_s-r^\phi_s\big )c^a(s,T) +\big (r^\phi_s-r^c_s\Big ) \alpha_s \eps_s+ (h_s- r_s) \tilde H_s\big ] ds \big |\cF_t\Big]
\end{aligned}
\end{equation}

\begin{Remark}

Following  \cite {BFP1}, a few issues about the above BSDE   need to be addressed.  

\begin{enumerate}

\item 
We remark that this equation has a unique strong solution as long as we take square integrable close-out value  and intensities and, for instance, we assume  the processes $r, r^c, r^\phi, h$ to be bounded. This is going to be our standing assumption.

\item The process $\tilde H_t$ is linked to the solution of the BSDE.
If we restrict to a diffusion setting with deterministic coefficients, the theory of BSDE's gives an explicit representation for the process $\tilde H$.  To deal with this, we extend the observation made in  \cite{BFP1}  when they assume deterministic intensities.

More precisely, we assume that the stock price, $S_u= \e^{X_u}$,  and the intensities processes, under the given risk-neutral probability, verify
$$
\begin{aligned}
dS_u =&r_uS_udu + \sigma(t, S_u) dY_u,\quad \text{and}\\
d\lam_u^i=& a_i(u, \lam_u^i)du + b_i(u, \lam_u^i)d B^i_u, \quad i=1,2,
\end{aligned}
$$
for correlated Brownian motions $Y, B^1, B^2$ and deterministic coefficients  $\sigma(u,x), a_i(u, \lam), b_i(u, \lam)$ chosen to ensure the existence and uniqueness  of strong solutions. Then 
 \eqref{BSDE1}  can be equivalently written on $\{\tau>t\}$ as
\begin{equation}
\label{BSDE2}
\begin{aligned}
&\e^{-\int_0^t (r_u+\lambda_u)du} c^a(t,T) = c^a(0,T)+\int_0^t Z_s dY_s + M_t \\
&-\int_0^t\e^{-\int_0^s (r_u+\lambda_u)du}  \big [\lambda_s \eps_s
- (1-\alpha_s)\Big (\lambda^1_s L_1(s)\eps_s^+ - \lambda^2_sL_2(s)\eps_s^- \big ) \Big ]ds\\
&- \int_0^t\e^{-\int_0^s (r_u+\lambda_u)du} \Big  [\big ( r_s-r^\phi_s\big )c^a(s,T) +\big (r^\phi_s-r^c_s\Big ) \alpha_s \eps_s+ (h_s- r_s) \tilde H_s\Big ] ds,
\end{aligned}
\end{equation}
where $Z$ is  the component of the solution of the BSDE coming from the martingale representation theorem, while  $M$ is a martingale  depending on the intensities and possibly on some  other stochastic factors (again represented by diffusions). In this context, 
 $c^a(t, T)$ is a deterministic function of the state variables, and assuming enough regularity of this function, $\tilde H$ should represent the $\delta-$hedging of the contract
$$
\tilde H_u=\frac{ \partial c^a(u,T)}{\partial S} S_u.
$$
On the other hand, the Markovian setting gives also that $Z$ is given by
$$
Z_u = \sigma(u,S_u)\frac{ \partial c^a(u,T)}{\partial S} \quad \Rightarrow\quad 
 \tilde H_u=\frac{ S_u}{\sigma(u, S_u) }Z_u,
$$
provided that $\sigma(u, x)>0$ for all $u,x$. 
\end{enumerate}
\end{Remark}

From now on, in addition to the hypotheses stated in the first of the previous remarks,  we assume that 
$$
0<\sigma_0 x \le \sigma(u,x)\le \sigma_1 x, \qquad \forall \, u,x
$$
for some constants $\sigma_0$ and $\sigma_1$.

This implies, as in \cite {BFP} or \cite {BFP1}, that we may apply Girsanov's theorem to change the Brownian motion driving the above BSDE  to include the term $\tilde H$. Indeed,
$$
B_t= Y_t  + \int_0^t (r_u - h_u) \frac {S_u}{\sigma(u, S_u)} du
$$ 
is a new Brownian motion with respect to the probability defined by the Radon-Nykodim derivative
$$
\frac {d\Q}{d\mathbb P}= \e^{ - \int_0^T  (r_u - h_u)  \frac {S_u}{\sigma(u, S_u) }dY_u+ \frac 12  \int_0^T  (r_u - h_u)^2  \frac {S^2_u}{\sigma^2(u, S_u) }du}
$$
which verifies the Novikov condition. Consequently, under $\Q$ the asset price equation and \eqref{BSDE2} become
\begin{equation}
\label{BSDE3}
\begin{aligned}
&dS_t = S_t h_tdt + \sigma(t, S_t) dB_t\\
&\e^{-\int_0^t (r_u+\lambda_u)du} c^a(t,T) = c^a(0,T)+\int_0^t Z_s dB_s + M_t \\
&-\int_0^t\e^{-\int_0^s (r_u+\lambda_u)du}  \big [\lambda_s \eps_s
- (1-\alpha_s)\big (\lambda^1_s L_1(s)\eps_s^+ - \lambda^2_sL_2(s)\eps_s^- \big ) \big ]ds\\
&- \int_0^t\e^{-\int_0^s (r_u+\lambda_u)du} \big  [\big ( r_s-r^\phi_s\big )c^a(s,T) +\big (r^\phi_s-r^c_s\Big ) \alpha_s \eps_s \big ] ds.
\end{aligned}
\end{equation}
Passing again to the conditional expectation and multiplying  both sides by $\e^{\int_0^t (r_u+\lambda_u)du}$,   we obtain
\begin{equation}
\label{BSDE4}
\begin{aligned}
&\mathbf 1_{\{\tau>t\}}c^a(t,T) =\mathbf 1_{\{\tau>t\}}\E_\Q\Big [
\e^{-\int_t^T( r_u+ \lambda_u)du} f(X_T) \\
&+\int_t^T\e^{-\int_t^s (r_u+\lambda_u)du}  \big [\lambda_s \eps_s
- (1-\alpha_s)\big (\lambda^1_s L_1(s)\eps_s^+ - \lambda^2_sL_2(s)\eps_s^- \big ) \big ]ds\\
&+ \int_t^T\e^{-\int_t^s (r_u+\lambda_u)du} \big  [\big ( r_s-r^\phi_s\big )c^a(s,T) +\big (r^\phi_s-r^c_s\Big ) \alpha_s \eps_s \big ] ds \big |\cF_t\Big].
\end{aligned}
\end{equation}
The latter equation  is linear or nonlinear depending on the choice of $\eps_s$. In the literature there are fundamentally two possible choices: either $\eps_s= c(s,T)$ (the default-free value of the claim) or  $\eps_s = c^a(s,T)$.

The first choice will always  give  a solvable linear BSDE. With the second choice, we might obtain a  solvable linear BSDE if the adjusted value stays always nonnegative  (or nonpositive), otherwise the negative and positive parts generate a nonlinear, not explicitly solvable, BSDE.

To exploit explicit formulas, when possible, we decide to choose  always $\eps_s= c(s,T)$ (that corresponds to asking collateralization proportional to the default-free price rather than to the current price), to guarantee the solvability of the BSDE for all European claims.

With this choice  \eqref {BSDE4}  becomes on $\{\tau>t\}$
$$
c^a(t,T) =\E_\Q\Big [
\e^{-\int_t^T( r_u+ \lambda_u)du} f(X_T) +\int_t^T\e^{-\int_t^s (r_u+\lambda_u)du}  \big [\Psi_s +( r_s-r^\phi_s\big )c^a(s,T) \big ] ds \big|\cF_t\Big]
$$
where 
$$
\Psi_s =\big[\lambda_s+(r^\phi_s-r^c_s)\alpha_s \big]  c(s,T)- (1-\alpha)\big [\lambda^1_s L_1(s)c(s,T)^+ - \lambda^2_s L_2(s)c(s,T)^- \big],
$$
which can be solved obtaining
\begin{equation}
\label{BSDE5}
\mathbf 1_{\{\tau>t\}} c^a(t,T)=\mathbf 1_{\{\tau>t\}} 
\E_\Q\Big [ \e^{- \int_t^T ( r^\phi_u+ \lambda_u)du} f(X_T)+\!\! \int_t^T \!\!\!
\e^{- \int_t^s (r^\phi_u+ \lambda_u)du} \Psi_s ds\big|\cF_t \Big].
\end{equation}
We remark we could have proposed a more general situation, considering  different collateral rates  and recovery processes and  close-out values for  the two parties. All these generalizations would have led to a more articulate, but not mathematically more difficult, equation. Indeed, the main nonlinearity is due to the recovery terms, once one decides to consider as close-out value the adjusted price of the contract.

In the next section, we introduce the market model and in the following two, we describe our evaluation procedure  by steps, leading to approximations  handier than Monte Carlo simulations. 

\begin{Remark}
We remark that if we are in absence of default of either part, $\lam^1=\lam^2=0$, funding, collateralization, rehypothecation are considered and the close-out value is taken equal to the contract's current value,  then the solution of (\ref{BSDE4}) becomes
$$
c^a (t,T)= \E_\Q\Big [ \e^{- \int_t^T [(1-\alpha_u)r^\phi_u+\alpha_u r^c_u]du} f(X_T)\big|\cF_t\Big ],
$$
which reduces to the usual Black \& Scholes setting, only if  the collateralization, funding, repo rates all coincide with the risk-free rate.
\end{Remark}

From now on we omit the probability $\Q$ in the notation of the expectation and we will  always  be referring to \eqref{BSDE5}.

\section{The evaluation procedure}

In what follows we specify the market model, where  the asset price  is represented as  a stochastic exponential, and the default intensities are assumed to be affine processes. Then  we illustrate a conditioning procedure that  helps to exploit explicit expressions for the default-free price, as it happens in the Black \& Scholes model when considering  European Vanilla Options or Futures. Finally, we apply a change of Numeraire that allows using the well-known expression for Zero-Coupon Bonds when interest rates are affine processes. This last step helps  to disentangle the contribution due to the intensities and the one coming from the derivative.

In  section \ref{val} we specialize this procedure to the case when the intensities are CIR  processes.
We will be able to derive semi-explicit formulas, that we approximate by means of a Taylor's expansion with respect to the correlation parameters,  up to the first or second order.  We do not consider the other very popular affine Vasicek model since it is well known explicit formulas can be derived in this case.

\subsection{The model}

We keep  denoting by $t\in [0, T]$ the initial time and  we make the following simplifying hypotheses
for \eqref{BSDE5}:

\begin{enumerate}
\item   all the rates, $r, r^c, r^\phi, h$ are deterministic;

\item for $ i=1,2$,  $(1-\alpha)L_i $ are constant and we will keep denoting them simply by $L_i$.
\end{enumerate}
So we have
$\displaystyle
\Psi_s=\big [\lambda_s+(r^\phi_s-r^c_s)\alpha \big]  c(s,T)-\big [\lambda^1_s L_1c(s,T)^+- \lambda^2_sL_2c(s,T)^-\big ].$
We also choose the following model for our state variables for fixed initial conditions $(t,x,\lambda_1, \lambda_2) \in \mathbb{R}^+\times \mathbb{R} \times \mathbb{R}^+ \times \mathbb{R}^+$, $\forall s \in [t,T]$
\begin{eqnarray}
\label{SDE1}
X_s\!\!\!\!& =&\!\!\!\! x + \int_t^s (h_u-\frac{\sigma^2}2)du + \sigma(B_s-B_t)\qquad x\in \mathbb R \\
\label{SDEi} \lam^i_s\!\!\!\!&=&\!\!\! \!
\lam_i +\!\! \int_t^s \!\![\gamma^i_u \lam^i_u+ \beta^i_u]du + \!\!\int_t^s\!\![ \eta^i_u\lam^i_u+ \delta^i_u]^{\frac 12} dB^i_u,\,\,\lam_i>0, \,\,i=1,2\!\!
\end{eqnarray}
where   $\sigma >0$ and $r,  \gamma^i, \beta^i,  \eta^i, \delta^i$, $i=1,2$ are all deterministic bounded functions of time, while
$(B^1,B^2,B^3)$ is a 3-dimensional Brownian motion, with 
$$
B_s= \rho_1 B^1_s+ \rho_2 B^2_s+ \sqrt{1-\rho_1^2- \rho_2^2} B^3_s, \qquad\rho_1^2+\rho_2^2 \le 1.
$$
The processes $X_s, \lambda^1_s, \lambda^2_s$ are Markovian, therefore $c(s,T)$ and  $c^a(s,T)$ are deterministic functions respectively of the state variables  $X$ and $(X, \lam^1,\lam^2)$, and depending also on the correlation parameters  $\brho=(\rho_1, \rho_2)$.

For any $t\le s\le T$, we define the processes
\begin{equation}
\label{bondlam0}
N_i(u,s):=\E(\e^{-\int_t^s\lam^i_v dv}|\cF_u),\quad i=1,2,
\end{equation}
which are martingales for $t\le u\le s$ and that, having chosen
the intensities as affine processes, by Fourier transform  have an explicit expression for their initial values
\begin{equation}
\label{bondlam}
N_i(t,s)=\e^{A_i(t,s)\lam_i+B_i(t,s)}\,  \Rightarrow \,N_i(u,s)=\e^{A_i(u,s)\lam_i+B_i(u,s)- \int_t^u \lam^i_vdv},
\end{equation}
where $\lam_i$ is the initial condition of the intensity  and $A_i$ and $B_i$ are deterministic functions
verifying a set of Riccati equations. 
We remark that by independence of the intensities we  also have
$$
N(u,s):=\E(\e^{-\int_t^s\lam_v dv}|\cF_u)=\E(\e^{-\int_t^s (\lam^1_v+\lam^2_v )dv}|\cF_u)=N_1(u,s)N_2(u,s),
$$
which  is still a martingale as product of independent martingales.
By   applying It\^o's formula, the dynamics of these martingales are given by
\begin{equation}
\label{dynbond}
\begin{aligned}
dN_i(u,s)&= N_i(u,s)A_i(u,s) (\eta^i_u\lam^i_u+ \delta^i_u)^{\frac 12}dB^i_u\\
dN(u,s)&= N(u,s)\Big [A_1(u,s) (\eta^1_u\lam^1_u+ \delta^1_u)^{\frac 12}dB^1_u +A_2(u,T) (\eta^2_u\lam^2_u+ \delta^2_u)^{\frac 12}dB^2_u\Big ].
\end{aligned}
\end{equation}
In some classical specifications of the  affine modeling framework:
\begin{itemize}
\item $\gamma^i_u=-\gamma_i$, \;$\beta^i(\lam)=\gamma_i \theta_i$,\; \;$\delta^i_u=\delta_i^2$,\;$\eta^i_u=0$\;\;(Vasicek)\\
\item $\gamma^i_u=-\gamma_i$, \;$\beta_i(\lam)=\gamma_i \theta_i$,\; $\delta^i_u=0$,\;$\eta^i_u=\eta_i^2$\;\;\;(CIR),
\end{itemize}
for $\gamma_i, \theta_i, i=1,2$ positive constants, 
 it is possible to compute $A_i(t,s)$ and $B_i(t,s)$ in closed form.

\subsection{Conditioning}

\label{cond}

In this subsection, we express  an alternative  formulation for the expectations in  \eqref{BSDE5}, which may be useful to write (conditionally) whenever possible,  the  explicit formula for the default-free price.
To simplify notation, from now on we denote by $\E_t$ the conditional expectation with respect to $\cF_t$.

 Since the interest rate $r^\phi$  is deterministic, we rewrite  \eqref{BSDE5}  as
\begin{equation}
\label{BSDE6}
\begin{aligned}
\mathbf 1_{\{\tau>t\}} c^a(t,T)=&\mathbf 1_{\{\tau>t\}} \Big \{
\e^{- \int_t^T r^\phi_u du} \E_t\Big (\e^{-\int_t^T\lambda_udu} f(X_T) \Big) \\
+&\mathbf 1_{\{\tau>t\}} \int_t^T \e^{- \int_t^s r^\phi_udu}\E_t\Big (\e^{- \int_t^s \lambda_udu} \Psi_s\Big)ds
\end{aligned}
\end{equation}
and we focus on the inner expectations.


\begin{Proposition}
Let 
$$
\cA^t_s=\cF^{B^1, B^2}_s\lor \cF_t= \sigma (\{B^1_u, B^2_u, u\le s\})\lor \cF_t,\quad t\le s\le T.
$$
Then 
 $$
\E_t\Big [\e^{-\int_t^T\lambda_udu} f(X_T) \Big]=\e^{\int_t^Th_udu}\E_t\Big [\e^{-\int_t^T\lambda_u du} \E\Big ( \e^{-\int_t^Th_udu} f(X_T) \Big |\cA^t_T\Big )\Big ],
$$
where $\ds X_T\Big |\cA^t_T\sim \cN\Big (\zeta_T(\brho)+  \int_t^T \big (h_udu -\frac{\Sigma^2(\brho)}2\big )du; \Sigma^2(\brho)(T-t)\Big )$ and 
$$
\zeta_T(\brho)= x +  \sigma(B^1_T-B^1_t)\rho_1+ \sigma(B^2_T-B^2_t) \rho_2-\frac{\sigma^2|\brho|^2}2(T-t), \quad
\Sigma(\brho)=\sigma\sqrt{1-|\brho|^2}.
$$
\end{Proposition}
\Proof From (\ref{SDE1}) the log-price at time $T$ is
$$
X_T=\zeta_T(\brho) + \int_t^T h_udu +\Sigma(\brho)(B^3_T-B^3_t)-\frac{\Sigma^2(\brho)}2(T-t),
$$
and a simple application of the  conditional expectation's tower-property  gives
$$
\begin{aligned}
\E_t\Big [\e^{-\int_t^T\lambda_udu} f(X_T) \Big]&=\E_t\Big [\E\Big (\e^{-\int_t^T\lambda_udu} f(X_T) \Big |\cA^t_T\Big )\Big ]=\E_t\Big [\e^{-\int_t^T\lambda_udu} \E\Big (f(X_T) \Big |\cA^t_T\Big )\Big ]\\
&=\e^{\int_t^Th_udu}\E_t\Big [\e^{-\int_t^T\lambda_u du} \E\Big ( \e^{-\int_t^Th_udu} f(X_T) \Big |\cA^t_T\Big )\Big ]. \Box
\end{aligned}
$$

\subsection{Changing Numeraires}
\label{secchange}

As a final step to evaluate the expectations $E_t$ in the previous expression,  we apply the following family of changes of probability
\begin{equation}
\label{changes}
\frac{d\Q^s}{d\Q}\Big|_{\cF_s}= \frac {N(s,s)}{N(t,s)},
\end{equation}
defining the $s-$forward measures, for  any $t\le s\le T$.
Recalling \eqref{dynbond},  by Girsanov's theorem, under  $\mathbb Q^s$
$$
W^i_v = B^i_v -\int_t^v A_i(u,s) (\eta^i_u\lam^i_u+ \delta^i_u)^{\frac 12}du, \quad i=1,2, \,t\le  v\le s
$$ 
define independent Brownian motions and the market dynamics, for $t\le v\le  s\le T$, become
\begin{eqnarray}
\hspace{-0.7cm}
\label{SDE2}
X_v\!\!\!\!& =&\!\!\!\! x + \int_t^v\Big (h_u-\frac{\sigma^2}2+\sigma\sum_{i=1,2} \rho_iA_i(u,s) (\eta^i_u\lam^i_u+ \delta^i_u)^{\frac 12}\Big)du + \sigma(W_s-W_t)\ \\
\hspace{-0.7cm}
\label{SDE2i} \lam^i_v\!\!\!\!&=&\!\!\! \!
\lam_i +\!\! \int_t^v \Big[(\gamma^i_u + A_i(u,s) \eta^i_u)\lam^i_u+ (\beta^i_u+ A_i(u,s)\delta^i_u)\Big]du + \!\!\int_t^v\!\![ \eta^i_u\lam^i_u+ \delta^i_u]^{\frac 12} dW^i_u ,
\end{eqnarray}
where   $(W^1,W^2,B^3)$ is a 3-dimensional Brownian motion, on $[t,s]$ with 
$$
W_v= \rho_1 W^1_v+ \rho_2 W^2_v+ \sqrt{1-\rho_1^2- \rho_2^2} B^3_v, \qquad\rho_1^2+\rho_2^2 \le 1
$$
and  we may conclude that the affine structure of the model is preserved. We remark that for each fixed $s$, different Brownian motions are generated. We keep denoting them in the same manner, as they all have the same distributional properties.

Hence, for any $t\le s\le T$ and any $\cF_s-$measurable random variable $Y $, we have  
\begin{equation}
\label{change}
\E_t\Big (\e^{\int_t^s\lambda_udu} Y\Big)=N(t,s)\E^s_t\big ( Y\big),
\end{equation}
where $\E^s_t$,  denotes  expectations  under $\Q^s$.

\section{Semiexplicit formulae}
\label{val}

In this section, we restrict to considering a European call with strike price $\e^\kappa$ and maturity $T$, for which we may exploit the Black \& Scholes formula, at least in a conditional fashion. We remark that in this case, by exploiting the  put-call parity, it is possible to extend the evaluation method also to forward contracts.

We treat the  case when the intensities are both described by a CIR  process. We do not consider here the Vasicek model,    since  not appropriate for intensities, as it does not guarantee the positivity of the process,  even though it has been previously considered in  credit risk modeling (see for instance \cite {Fard15}) as it allows to write very computable explicit formulas.

\subsection{The CIR specification}

In this case, the dynamics of the market, for any $t\le s\le T$,  are given by

\begin{eqnarray}
\hspace{-0.7cm}
\label{CIR0}
X_s\!\!\!\!& =&\!\!\!\! x + \int_t^s\Big (h_u-\frac{\sigma^2}2\Big )du+\sigma(B_s-B_t)\ \\
\hspace{-0.7cm}
\label{CIRi} \lam^i_s\!\!\!\!&=&\!\!\! \!
\lam_i +\!\! \int_t^s\gamma_i(\theta_i-\lam^i_u)du + \eta_i\int_t^s\!\ \sqrt{\lam^i_u} dB^i_u,\quad i=1,2.
\end{eqnarray}
We denote by $\tilde r_u = r^\phi_u - h_u$ and , $\hat r_u=r^\phi_u - r^c_u$
we have to compute
\begin{equation}
\label{Boh}
\begin{aligned}
\mathbf 1_{\{\tau>t\}} c^a(t,T)&=\mathbf 1_{\{\tau>t\}} \left \{\e^{- \int_t^T \tilde r_udu}
\E_t\Big [ \e^{- \int_t^T \lambda_u du}  \e^{- \int_t^T h_udu}f(X_T) \Big]\right .\\
&\left .+\!\! \int_t^T \!\!\!\e^{- \int_t^s r^\phi_u du}\E_t\Big [ 
\e^{- \int_t^s\lambda_udu} \Lambda_s c(s, T)\Big]ds\right \}
\end{aligned}
\end{equation}
where
$$
\Lambda_s= \lam_s+ \alpha\hat r_s -L_1 \lam^1_s.
$$

\begin{Proposition}
Let $f(x)= (\e^x-\e^\kappa)^+$ and
$$
\begin{aligned}
c(s, T)&\equiv c(s, T)^+=  c_{BS} (X_s,s,\bar v_s,\sigma)\\
c_{BS} (x,s,\bar v _s,\sigma)&=\e^x \cN\big (d_1(x,s,\bar v_s , \sigma)\big )-
\e^{\kappa - \bar v_s} \cN\big (d_2(x,s,\bar v_s, \sigma)\big )\\
d_{1,2}(x, s, \bar v_s, \sigma)&= \frac { x- \kappa+\bar v_s\pm 
\frac{\sigma^2}2 (T-s)} { \sigma\sqrt{(T-s) }},
\end{aligned}
$$
where we denoted by
$\ds \bar v_s=\int_s^T v_udu $, for any $v: [0,T] \longrightarrow \mathbb R$. Then we have
\begin{equation} 
\label{Boh1}
\begin{aligned}
\mathbf 1_{\{\tau>t\}} c^a(t,T; \brho)&=\mathbf 1_{\{\tau>t\}} \left\{\e^{- \int_t^T \tilde r_udu}
\E_t\Big [ \e^{- \int_t^T \lambda_u du}  c_{BS}\big (\zeta_T(\brho), t, \bar h_t, \Sigma(\brho)\big )\Big]\right .\\
&+\left .\!\! \int_t^T \!\!\!\e^{- \int_t^s r^\phi_u du}\E_t\Big [\e^{- \int_t^s\lambda_udu} \Lambda_s c_{BS}\big (X_s(\brho), s,  \bar r_s, \sigma\big )\Big ]ds\right \}.
\end{aligned}
\end{equation}
\end{Proposition}

\Proof Applying inside the first expectation the conditioning with respect to $\cA^t_T$,  we obtain 
$$
\begin{aligned}
&\E_t\Big [ \e^{- \int_t^T \lambda_u du}  \e^{- \int_t^T h_udu}f(X_T) \Big]=
\E_t\Big [\E_t\Big ( \e^{- \int_t^T \lambda_u du}  \e^{- \int_t^T h_udu}f(X_T)|\cA^t_T\Big ) \Big]\\
=& \E_t\Big [ \e^{- \int_t^T \lambda_u du}  \E_t\Big ( \e^{- \int_t^T h_udu}f(X_T)\Big|\cA^t_T\Big  ) \Big]=
 \E_t\Big [ \e^{- \int_t^T \lambda_u du}  c_{BS}\big (\zeta^t_T(\brho), t, \bar h_t, \Sigma(\brho)\big )\Big]
\end{aligned}
$$
and we may view the second  expectation in \eqref{Boh} as $\displaystyle \E_t\Big [ \e^{- \int_t^s\lambda_udu} \Lambda_sc_{BS} (X_s(\brho),s,\bar r_s, \sigma) \Big] $ where ,for $t\le s\le T$, setting   $M^i_s= B^i_s-B^i_t $, for $i=1,2$, we have 
$$
X_s(\brho)=x + \int_t^s\Big (h_u-\frac{\sigma^2}2\Big )du+\sigma\big (M^1_s\rho_1+M^2_s \rho_2 + M^3_s \sqrt{1 - |\brho|^2}\big ).
$$
Consequently,  we have 
$$
d_{1,2}(\zeta_T(\brho), t,\bar h_t,\Sigma(\brho))=\begin{cases}& \left [
d_1(x,t,\bar h_t, \sigma)+ \frac{M^1_T}{\sigma\sqrt{T-t}}\rho_1+\frac{M^2_T}{\sigma\sqrt{T-t}}\rho_2- \sigma \sqrt{T-t}|\brho|^2\right ]
\frac 1 {\sqrt{1-|\brho|^2}}\\
& \Big [d_2(x,s,\bar h_t,\sigma)+ \frac{M^1_T}{\sigma\sqrt{T-t}}\rho_1 +\frac{M^2_T}{\sigma\sqrt{T-t}}\rho_2\Big ]\frac 1 {\sqrt{1-|\brho|^2}}.
\end{cases}
$$
Pointing out the dependence on $\brho$ of $c^a(t,T)$, we get \eqref{Boh1}. $\Box$

\medskip

We want to approximate (\ref{Boh1}) by a  Taylor expansion with respect to the correlation parameters $\brho=(\rho_1, \rho_2)$ around $\bm 0= (0,0)$ on $\{\tau>t\}$. The first-order approximation would hence be
$$
c^a(t,T; \brho)\approx c^a(t,T; \bm 0)+ \frac {\partial c^a(t,T; \bm 0)}{\partial \rho_1}\rho_1
+ \frac {\partial c^a(t,T; \bm 0)}{\partial \rho_2}\rho_2.
$$
\begin{Remark}
 For the sake of exposition, we decided to restrict our discussion to the first order approximation, which may turn to be   extremely satisfying when the model seems to exhibit a roughly linear dependence upon the correlation parameters. This was highlighted by the  Monte Carlo simulations for the CIR intensity setting (section \ref{num}) and the accuracy of our method turned out to be very good.
If the dependence on the correlation parameters is more markedly nonlinear,  one may develop Taylor's polynomial to  a higher order to capture this behavior. We explicitly wrote also a second-order formula: it is  computationally longer, but it does not present any additional theoretical complexity. We did not report it here
to keep the  exposition lighter.

\end{Remark}
Since the integrability conditions are satisfied, the derivatives pass under the integral and expectation signs and the problem is reduced to computing the derivatives with respect to the correlation parameters of  $c_{BS}\big (\zeta_T(\brho), t, T, \Sigma(\brho)\big )$ and of $c_{BS}\big (X_s(\brho), s, T, \sigma\big )$ and evaluating them at $\bm 0$.
After some calculations, one arrives at the  following expressions 
$$
c_{BS}\big (\zeta_T(\brho), t,\bar h_t, \Sigma(\brho)\big )\approx  c_{BS}\big (x, t, \bar h_t, \sigma\big )+
\sigma \e^x\cN\big (d_1(x, t,\bar h_t, \sigma)\big )\!\!
\left [ M^1_T\rho_1 + M^2_T\rho_2 \right ]
$$
and 
$$
c_{BS}\big (X_s(\brho), s, \bar r_s, \sigma\big )\approx \, c_{BS}\big (X_s(\bm 0), s,\bar r_s,  \sigma\big )+
\sigma \e^{X_s(\bm 0)}\cN\big (d_1(X_s(\bm 0), s,\bar r_s, \sigma)\big )\!
\left [ M^1_s\rho_1 + M^2_s\rho_2 \right ]
$$
to be plugged into \eqref{Boh}, with each term to be computed following the procedure outlined in the previous sections. 
Thus,  exploiting the independence between $X_s(\bm 0)$ and $B^1, B^2$ we have
$$
\label{repre}
\begin{aligned}
c^a(t,T; \brho)\approx &
\e^{- \int_t^T \!\!\tilde r_udu}\!
\Bigg\{\!N(t, T)c_{BS}\big (x, t, \bar h_t,\sigma\big )
+\!\sigma \e^x\cN\big (d_1(x, t,\bar h_t, \sigma)\big)\E_t\Big [\e^{- \int_t^T \!\!\lambda_u du}(M^1_T \rho_1\!+\!M^2_T \rho_2)\Big ]\Bigg \}\\
& 
+\!\!\int_t^T\!\!\! \e^{- \int_t^s r^\phi_udu}\Bigg \{\E_t\Big [ \e^{\int_t^s\lam_u du}\Lambda_s\Big ]\E_t\Big [c_{BS}\big (X_s(\bm 0), s, \bar r_s,\sigma\big )\Big ] 
\\
&\hspace{2.4cm}+\sigma\E_t\Big [\e^{X_s(\bm 0)}\cN\big (d_1(X_s(\bm 0), s,\bar r_s, \sigma)\big )\Big ]\sum_{i=1}^2\E_t\Big (\e^{- \int_t^s \lambda_u du}\Lambda_sM^i_s\Big )\rho_i 
 \Bigg \}ds
\end{aligned}
$$
and we have to compute every single expectation. We proceed by steps, showing that we may reduce to computing some basic cases.

\begin{enumerate}

\item
Noticing that 
$$
\begin{aligned}
&M^3_s\sim N(0; \sigma^2 (s-t))\\
&X_s(\bm 0)= x+\int_t^s (h_u-\frac{\sigma^2}2)du+ M^3_s \sim N \Big (x+\int_t^s (h_u-\frac {\sigma^2}2) du; \sigma^2 (s-t)\Big ),\\
&d_i(X_s(\bm 0), s,\bar r_s,\sigma)=\frac {X_s(\bm 0)\!-\!k \!+\bar r_s \pm \!\frac {\sigma^2}2(T-s)}{\sigma\sqrt{T\!-\!s}} \\
&\hspace{2.9cm} =\frac {M^3_s}{\sqrt{T\!-\!s}}\!+\! d_i(x,s, \bar r_s,\sigma) + \frac{1}{\sigma \sqrt{T\!-\!s}} \int_t^s (h_u-\frac {\sigma^2}2) du\\
&\hspace{2.9cm}\sim  \, N\Big (d_i(x,s, \bar r_s,\sigma)+\frac{1}{\sigma \sqrt{T\!-\!s}} \int_t^s (h_u-\frac {\sigma^2}2) du, \frac{s-t}{T-s} \Big ), \ \ \ i=1,2\\
&\E_t\Big [c_{BS}\big (X_s(\bm 0), s, \bar r_s,\sigma\big )\Big ]=\E_t\Big [\e^{X_s(\bm 0)}\cN\big (d_1(X_s(\bm 0), s,\bar r_s, \sigma)\big )\Big ]
- \e^{\kappa- \bar r_s}\E_t\Big [\cN\big (d_2(X_s(\bm 0), s,\bar r_s, \sigma)\big )\Big ],
\end{aligned}
$$ 
the  Gaussian integrals can be computed explicitly 
$$
\begin{aligned}
\E_t\Big [\e^{X_s(\bm 0)}\cN\big (d_1(X_s(\bm 0), s,\bar r_s, \sigma)\big )\Big ]=&
\e^{x+\int_t^s h_udu}\cN\Big(d_1(x+(\bar r_s- \bar h_s),t,\bar h_t,\sigma)\Big )\\
\E_t\Big [\cN\big (d_2(X_s(\bm 0), s,\bar r_s, \sigma)\big )\Big ]=&
\cN\Big(d_2(x +(\bar r_s- \bar h_s),t,\bar h_t,\sigma) \Big ),
\end{aligned}
$$
by applying the following

\begin{Lemma} \label{zack}Let $p\in \mathbb\R$ and $X\sim N(\mu,\nu^2)$, then
$$
\E(\e^{pX}\cN(X))=\e^{p\mu+\frac{(p\nu)^2}{2}}\cN\biggl(\frac{\mu+p\nu^2}{\sqrt{1+\nu^2}}\biggr)
$$
where by $\cN$ we denote the standard Normal distribution function.
\end{Lemma}
\begin{Proof}
 see Zacks (1981) for $p=0$, the general case follows by a ``completing the squares" argument.\qquad $\square$
\end{Proof}

Therefore we may conclude that
\begin{equation} \label{cBS_h}
\E_t\Big [c_{BS}\big (X_s(\bm 0), s, \bar r_s,\sigma\big )\Big ]=\e^{-(\bar r_s- \bar h_s)+\int_t^s h_udu}
c_{BS}\big (x +(\bar r_s- \bar h_s), t, \bar h_t,\sigma\big )
\end{equation}
\item  It remains to evaluate  the expectations
$$
\E_t\Big (\e^{- \int_t^s \lambda_u du} \Lambda_s\Big), \quad 
\E_t\Big (\e^{- \int_t^s \lambda_u du} (B^i_s- B^i_t )\Big ), \quad 
\E_t\Big (\e^{- \int_t^s \lambda_u du}\Lambda_s\  (B^i_s- B^i_t)\Big )\quad i=1,2
$$
Recalling that $\Lambda_s= \lam_s+ \alpha\hat r_s -L_1 \lam^1_s$ the above expressions reduce to computing
$$
\E_t\Big [\e^{- \int_t^s \lambda_u du}(\lambda^i_s)^\alpha (B^j_s- B^j_t)^k \Big ]
$$
for $i,j=1,2$, and $\alpha, k=0,1$.
$$
\E_t\Big [\e^{- \int_t^s \lambda_u du}(\lambda^i_s)^\alpha (B^j_s- B^j_t)^k \Big ].
$$
To do so, we apply the change of Numeraire  described in   subsection \ref{secchange}, obtaining
$$
\E_t\Big [\e^{- \int_t^s \lambda_u du}(\lambda^i_s)^\alpha (B^j_s- B^j_t)^k \Big ]\!=\!
N(t, s)\E^s_t\Big [(\lambda^i_s)^\alpha \big[(W^j_s- W^j_t) +\eta_j\!\!\int_t^s \!\!\!A_j(u,s) \sqrt {\lam^j_u}du \big]^k\Big ].
$$
We can exploit the 
 independence of $W^1$ and $W^2$,  so that the last expectation, for $i\neq j$ becomes
$$
\eta_j\E^s_t\Big [(\lambda^i_s)^\alpha \Big ]\Bigg [\int_t^s A_j(u,s)\E^s_t\Big (\sqrt {\lam^j_u}\Big )du\Bigg]^k, 
$$
where for $t\le u\le s$
$$
 \lam^i_u=
\lam_i +\!\! \int_t^u \left [ \gamma_i\theta_i- \Big (\gamma_i-\eta_i^2A_i(v,s)\Big ) \lam^i_u\right ]dv + \eta_i\int_t^u\!\ \sqrt{\lam^i_v} dW^i_v.
$$
When  $i=j$,  if $k=0$, clearly we have only the first expectation, if $\alpha=0$ only the second, and  for  $\alpha=k= 1$, we end up with 
$$
\E^s_t\Big [\lambda^i_s (W^i_s- W^i_t) \Big ]+\eta_j\!\!\int_t^s \!\!\!A_i(u,s)\E^s_t\Big [\lambda^i_s \sqrt {\lam^i_u} \Big]du.
$$

\item
Thus we have  reduced the problem to considering the expectations for $u\le s$
\begin{eqnarray}
\label{ex1}
\!\!\!&&\!\!\!\E^s_t\big (\lambda^i_s \big ),\quad \E^s_t\big (\sqrt {\lam^i_u}\big ),\quad
\E^s_t\big (\lam^i_s\sqrt {\lam^i_u}\big ), \\
\label{ex2}
\!\!\!&&\!\!\!\E^s_t\big (\lam^i_s(W^i_s- W^i_t)\big ),
\end{eqnarray}
The third of \eqref{ex1}, again by the independence of the increments, can be written as
$$
\E^s_t\big (\lam^i_s\sqrt {\lam^i_u}\big )=\E^s_t\big ((\lam^i_s-\lam_u^i)\sqrt {\lam^i_u}\big )+\E^s_t\big ((\lam^i_u)^{\frac 32}\big )= \E^s_t\big (\lam^i_s-\lam_u^i\big  )\E^s_t\big (\sqrt {\lam^i_u}\big )+\E^s_t\big ((\lam^i_u)^{\frac 32}\big ).
$$
By applying It\^o's formula and taking expectations, for $t\le u\le s\le T$ we have 
$$
\begin{aligned}
\E^s_t\big (\lambda^i_u \big )\!&=\e^{-\int_t^u[\gamma_i- \eta_i^2A_i(\xi,s)]d\xi}
\left \{\lam_i+\gamma_i\theta_i\int_t^u \e^{\int_t^v[\gamma_i- \eta_i^2A_i(\xi,s)]d\xi}dv\right \},\\
\E^s_t\Big [\sqrt{\lambda^i_u} \Big ]\!&=\e^{-\frac 12\!\int_t^u[\gamma_i- \eta_i^2A_i(\xi,s)]d\xi}
\! \left [\!\sqrt{\lam_i} \!+\!\frac 12\Big[\gamma_i\theta_i\!-\!\frac{\eta_i^2}4\Big]\!\!\int_t^u \!\!\!\e^{\frac 12\!\int_t^v[\gamma_i- \eta_i^2A_i(\xi,s)]d\xi}\E^s_t\Big [\frac 1{\sqrt{\lambda^i_v} }\Big ]dv\right ]\!,\\
\E^s_t\Big [(\lambda^i_u)^{\frac 32} \Big ]\!&=\e^{-\frac 32\!\int_t^u[\gamma_i- \eta_i^2A_i(\xi,s)]d\xi}
\! \left [\!(\lam_i)^{\frac 32} \!+\!\frac 32\Big[\gamma_i\theta_i\!+\!\frac{\eta_i^2}4\Big]\!\!\int_t^u \!\!\!\e^{\frac 32\!\int_t^v[\gamma_i- \eta_i^2A_i(\xi,s)]d\xi}\E^s_t\Big [\sqrt{\lambda^i_v} \Big ]dv\right ],
\end{aligned}
$$
and we approximate $\frac 1{\sqrt{\lambda^i_v} }$ by $\displaystyle \frac 1{\sqrt{\lambda_i} }$ or $\displaystyle \frac 1{\sqrt{\theta_i} }$, freezing  the process either at the initial condition or at the mean reversion parameter. This choice usually provides simple  and numerically quite accurate approximations of the powers of a CIR process.
Finally, we may use integration by parts for the expectation  \eqref{ex2} and we may conclude 
$$
\E^s_t\big (\lam^i_s(W^i_s- W^i_t)\big )=\eta_i \int_t^s \e^{-\int_u^s[\gamma_i- \eta_i^2A_i(\xi,s)]d\xi}
\E^s_t\Big [\sqrt{\lambda^i_u} \Big ]du
$$
In conclusion, all the pieces appearing in  \eqref{repre} can be computed explicitly, provided we perform the mentioned freezing for $(\lambda^i_u)^{-\frac 12}$. 
\end{enumerate}
Summarizing
\begin{equation}
\label{firstapp}
c^a(t,T; \brho)\approx g_0(t,T; \bm 0)+ g_1(t,T; \bm 0)\rho_1+
g_2(t,T;\bm 0)\rho_2
\end{equation}
where the zeroth term is  (with $R_1= 1-L_1$)
\begin{equation}
\label{g0}
\begin{aligned}
&g_0(t,T; \bm 0)=
\e^{- \int_t^T \!\!\tilde r_udu}\!N(t, T) c_{BS}\big (x, t, \bar h_t, \!\sigma\big )\\
+&\int_t^T\!\!\! \e^{- \int_t^s \!  \tilde  r_udu-(\bar r_s-\bar h_s)}N(t,s)\big [R_1 \E^s_t(\lam^1_s)\!+\!\E^s_t(\lam_s^2)+\alpha \hat r_s\big]
c_{BS}\big (x +(\bar r_s\!-\!\bar h_s), t, \bar h_t,\sigma\big )ds\!
\end{aligned}
\end{equation}
and the first-order coefficients are
\begin{eqnarray}
\label{g1}
\begin{aligned}
&g_1(t,T; \bm 0)
 =\sigma \Bigg \{\eta_1 
 \e^{x- \int_t^T \!\!\tilde r_udu}\!N(t, T)\cN\big (d_1(x, t, \bar h_t,\sigma)\big ) \int_t^T \!\!A_1(s,T)\E^T_t\big (\sqrt {\lam^1_s}\big )ds
\\
&
+\!\!\int_t^T\!\!\! \e^{x- \int_t^s \tilde r_udu}N(t,s)\cN\Big(d_1(x+(\bar r_s- \bar h_s),t,\bar h_t,\sigma)\Big )
\Big [  R_1
\E^s_t\big (\lam^1_s(W^1_s\!\!-\!W^1_t)\big )\\
&+\eta_1\!\int_t^s \!\!A_1(u,s)
\big [
\E^s_t(\lam^1_s-\lam_u^1)\E^s_t(\sqrt {\lam^1_u} )+\E^s_t\big ((\lam^1_u)^{\frac 32}\big )+ \big (\E^s_t (\lam^2_s)+\alpha \hat r_s\big )\E^s_t(\sqrt {\lam^1_u}) 
\big ]du 
\Big]ds
\Bigg \}
\end{aligned}\\
\label{g2}
\begin{aligned}
&g_2(t,T; \bm 0)  =\sigma 
\Bigg\{
\eta_2 \e^{x- \int_t^T \!\!\tilde r_udu}\!N(t, T) \cN\big (d_1(x, t, \bar h_t,\sigma)\big ) \int_t^T \!\!A_2(s,T)\E^T_t\big (\sqrt {\lam^2_s}\big )ds
\\
&+ \!\!\int_t^T\!\!\! \e^{x- \int_t^s  \tilde r_udu}N(t,s)\cN\Big(d_1(x+(\bar r_s- \bar h_s),t,\bar h_t,\sigma)\Big )
\Big[
 \E^s_t\big (\lam^2_s(W^2_s\!\!-\!W^2_t)\big )\\
 &+\!\eta_2\! \!\int_t^s \!\!\!A_2(u,s)\big [\big( R_1\E^s_t (\lam^1_s)+\alpha\hat  r_s\big )\E^s_t(\sqrt {\lam^2_u})+
  \E^s_t(\lam^2_s-\lam_u^2)\E^s_t(\sqrt {\lam^2_u} )+\E^s_t\big ((\lam^2_u)^{\frac 32}\big )\big ]du 
\Big ]ds\!\!
\Bigg\}
\end{aligned}
\end{eqnarray}
where, for $t\le s\le T$ and $i=1,2$, we have
$$
N_i(t,s)=\e^{A_i(t,s)\lam_i + B_i(t,s)}, \quad N(t,s)= N_1(t,s)N_2(t,s)
$$
with
\begin{eqnarray*}
h_i= \sqrt{\gamma_i^2+2\eta_i^2} , \qquad A_i(t,T)&=&-\frac{2 (\e^{h_i (T-t)}-1)}{h_i-\gamma_i+(h_i+\gamma_i)\e^{h_i(T-t)}}\\
B_i(t,T)&=&\frac {2\gamma_i\theta_i}{\eta_i^2}
\ln\left( 
\frac{2h_i\e^{\gamma_i+h_i(T-t)}}{h_i-\gamma_i+(h_i+\gamma_i)\e^{h_i (T-t)}}
\right ).
\end{eqnarray*}

\section{Numerical results}

\label{num}

In this section, we present some numerical results of our approximation method for the call price.   As a first step, we assess the performance of the first-order approximation \eqref{firstapp} by using the Monte Carlo evaluations with control variates  as a benchmark, employing the default-free price as control: in the considered cases, this reduces the length of the confidence interval by at least one order of magnitude. For the simulations, we generated $M=10^6$ sample paths with a time step equal to $10^{-3}$ for any considered maturity. The benchmark Monte Carlo method was implemented to approximate the call price (\ref{BSDE5}) by using the Euler discretization scheme with full truncation for the intensity  processes $\lam_t^1$ and  $\lam_t^2$ (see \cite{LKVD}) and with an exact simulation of the Brownian motion for the underlying $X_t$. The running integrals appearing in the expectations were evaluated by means of trapezoidal routine. All the algorithms were implemented in MatLab (R2019b). 

The evaluation of the zeroth and first-order terms of our approximation (\eqref{g0}, \eqref{g1}, \eqref{g2}) requires the computation of nested one-dimensional integrals of well-behaved functions once for each set of chosen parameters and this step was implemented through the vectorized global adaptive quadrature MatLab algorithm.
 
The parameters of the intensity processes were set as in \cite{BRH18} and \cite{ARS2} (see Table \ref{Tab1}) and they agree with calibrated default intensities.  The strike price was fixed to $K=\e^{\kappa}=100$ and we considered two maturities, $T=0.5$ and $T=2$.
Lastly, without loss of generality, we took  $t = 0$, the log-asset's initial value was set to $4.6052$, and its volatility to $\sigma=40\%$. The remaining parameters were chosen as $r=h=0.001$, $r^{\phi}=0.005$, $r^{c}=0.002$ and $\alpha=0.5$.

\begin{figure}
\vspace{-5cm}
\hspace{-2.5cm}
\begin{subfigure}[b]{.6 \textwidth}
\includegraphics[width=12cm,height=17cm]{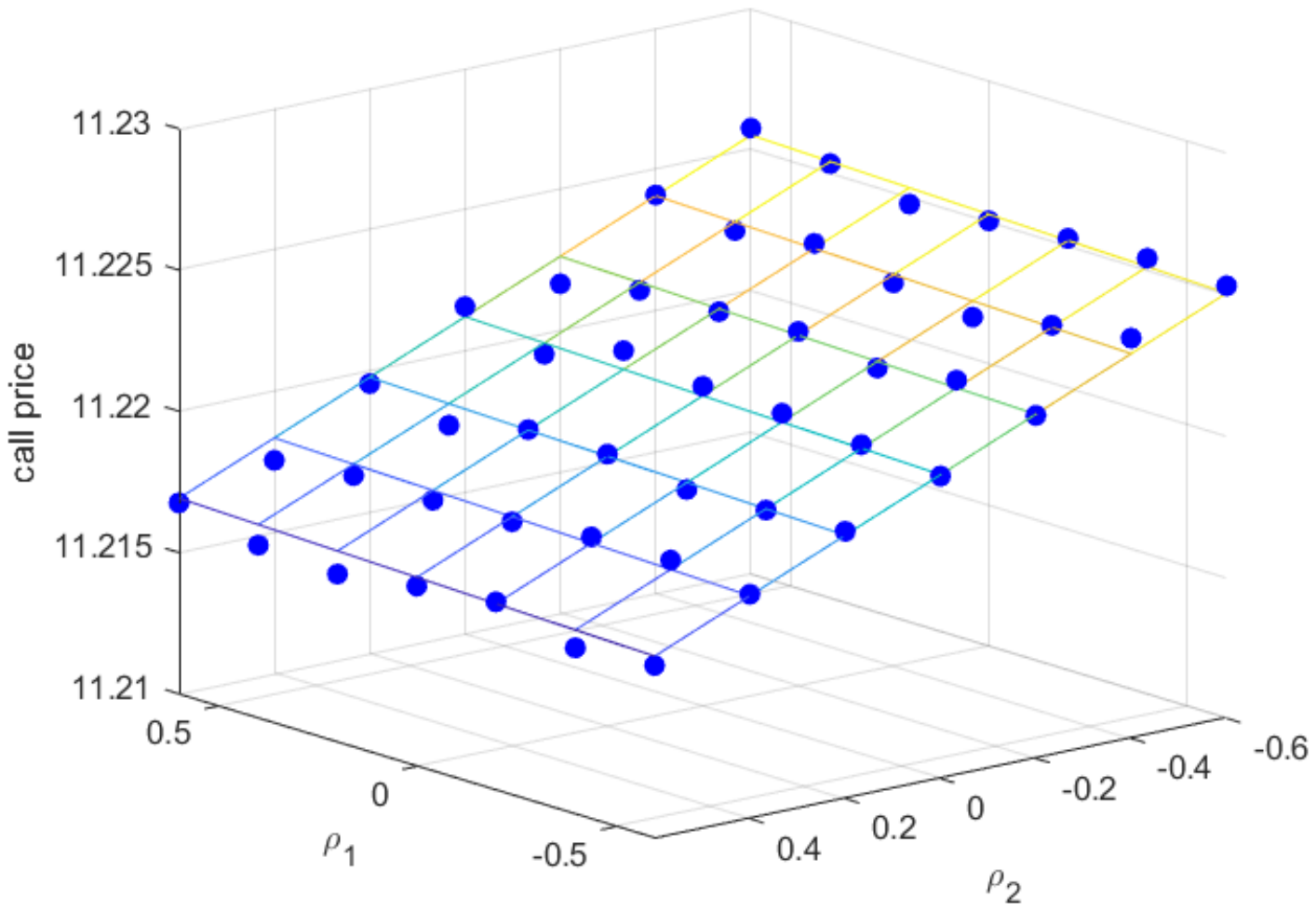}
\end{subfigure}
\hspace{-1cm}
\begin{subfigure}[b]{.6 \textwidth}
\includegraphics[width=12cm,height=17cm]{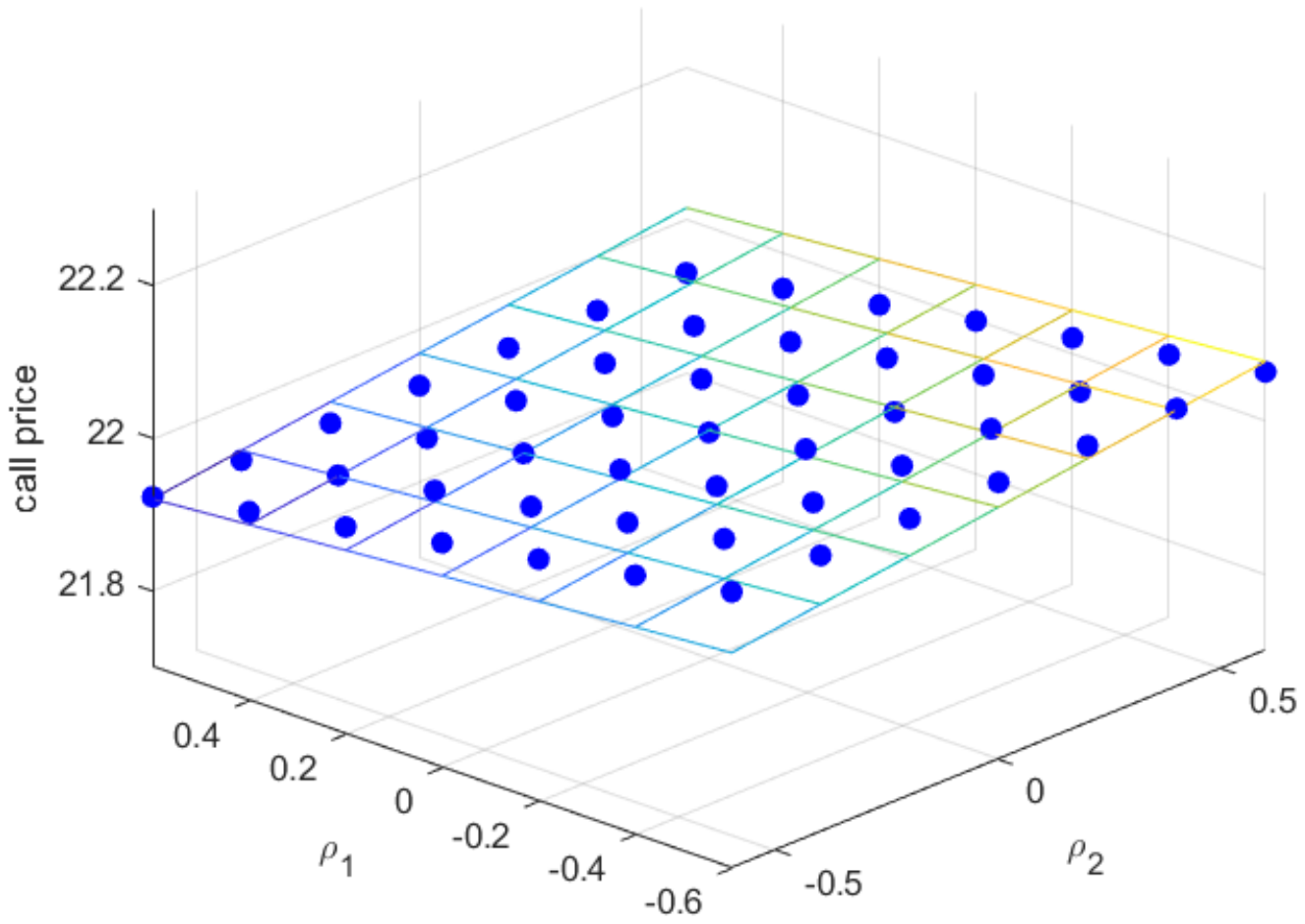}
\end{subfigure}
\vspace{-5.0cm}
\caption{\textit{MC prices (dot) vs approximated prices (lines).On the left $T=0.5$, on the right $T=2$.}}
\label{fig1}
\end{figure}

The accuracy of the first-order  approximation is summarized in Tables (\ref{Tab2}), (\ref{Tab3}), listing the errors with respect to the benchmark MC prices (see also figure (\ref{fig1})) for different choices of the default parameters for the Investor and the Counterparty and to the time-to-maturity $T$ of the contract. It is apparent how the approximation is highly satisfactory for short term maturity while it tends to deteriorate a little when the horizon increases.

In Table (\ref{Tab4}) we highlight the separate contributions of  the zeroth and first-order terms, $g_0(0,T; \bm 0)$, $g_1(0,T; \bm 0)$ and $g_2(0,T; \bm 0)$ in  (\ref{firstapp}), which are not significantly affected in  relative magnitude  by changes in the values of the  parameters. In particular, the  contribution due to the correlation between the underlying and  the  intensities is quite sizeable and it supports the choice of stochastic processes versus deterministic functions to represent the intensities. We notice that the contribution of the term $g_1$ is  more significant compared to that of $g_2$ which appears to be always rather small. This is to be expected since we are considering a call option and default of the Investor is bound to have a  limited impact on the overall value; on the contrary, the term $g_1$ is more relevant being connected to the counterparty's default and, as natural,  it decreases as the collateralization tends to one.

The contribution coming from the stochastic nature of the intensities can be better appreciated by looking at the results of the  further set of numerical experiments reported in Table (\ref{Tab5}). There, in order to compare with the results in   \cite{BFP1}, we considered the rates $r=0.001$, $h=0.005$, $r^{\phi}=0.005$, $r^{c}=0.002$ and we chose $\lambda_0^1=0.04$, $\lambda_0^2=0.02$  and the other parameters as in \eqref{Tab1}. The losses given default were set to $L_1=L_2=60\%$ and we took $T=0.5$. The correction that we obtain with respect to the prices in \cite{BFP1} is of the order of $10^{-2}$, which can, of course, become very relevant as the volume of the transaction grows.

As a final remark, we write explicitly our evaluation formula when 
 constant intensities $\lambda^i_t \equiv \lambda^i$ are taken. It is immediately seen by using (\ref{cBS_h}) that the price (\ref{Boh}) becomes
\begin{eqnarray} 
c^a(t,T) & = & \e^{(\lambda^1+\lambda^2 - (r^{\phi}-h)) (T-t)} c_{BS}(x,t,\bar h,\sigma) + (\lambda^1+\lambda^2 + (r^{\phi}-r^c)\alpha - \lambda^1 L_1) \times \nonumber \\
             &  &  \int_t^T \e^{-(\lambda^1+\lambda^2 + (r^{\phi}-h))(s-t)} \e^{-(r-h)(T-s)} c_{BS}(x+(r-h)(T-s),t,\bar h,\sigma) ds \label{ca_const}
\end{eqnarray}
which, as noticed in \cite{BFP}  and \cite{BFP1}, shows that  the interplay among all the rates in this framework accounts for a significant contribution to the global price.

Last but not least, we would like to  point out that our approximation implies a very big reduction of the computational time as it allows avoiding the costly multi-dimensional  Monte Carlo Simulations or PDE discretization.

\begin{table}[t]
\centering
\begin{tabular}{c|c|c|c|c|c|c}
  \hline
  & $\lambda_0$ & $\gamma$  & $\theta$ &  $\eta$ & 6-months surv. prob. & 2-years surv. prob.\\ \hline
$\tau_1$ (counterparty)  & 0.03  & 0.02  & 0.161 & 0.08  & 0.9848  & 0.9371 \\
$\tau_2$ (investor) & 0.035 & 0.35 & 0.45   & 0.15  & 0.9660  & 0.7399\\
\hline
\end{tabular}
\caption{Parameter sets for the CIR default intensities.}
\label{Tab1}
\end{table}

\begin{table}[t]
\centering
\begin{tabular}{c|ccccccc}
\hline
  & -0.6 & -0.4 & -0.2 & 0 & 0.2 & 0.4 & 0.6 \\ \hline
-0.6 &  -7.478e-04 &  -5.850e-04 &  -3.852e-04 &  -1.951e-04 & -4.362e-05  &  7.122e-05 &   1.881e-04 \\
-0.4 &  -5.338e-04 &  -3.423e-04 &  -1.508e-04 &   5.306e-05  &  1.955e-04  &  3.118e-04  &  3.636e-04 \\
-0.2 &  -3.104e-04 &  -9.415e-05 &   8.240e-05 &   2.456e-04 &   3.640e-04 &   4.693e-04 &   5.321e-04 \\
0     & -1.194e-04  &  8.440e-05 &   2.489e-04 &   4.203e-04 &   5.234e-04 &   6.252e-04 &   7.105e-04 \\
0.2 &   5.723e-05 &   2.527e-04 &   4.217e-04 &   5.816e-04  &  7.102e-04  &  8.091e-04  &  9.161e-04 \\
0.4  &  2.584e-04 &   4.708e-04 &   6.296e-04 &   7.458e-04 &   8.760e-04 &   9.736e-04 &   1.079e-03 \\
0.6 &   4.854e-04  &  6.768e-04  &  8.431e-04  &  9.614e-04 &  1.074e-03  &  1.167e-03 &  1.241e-03 \\ \hline
\end{tabular}
\caption{Approximation errors, Set 1 for $\tau_1$, Set 2 for $\tau_2$, $T=0.5$. The average length of the $95\%$ confidence interval for the MC estimates is $5.3939e-04$.}
\label{Tab2}
\end{table}

\begin{table}[t]
\centering
\begin{tabular}{c|ccccccc}
\hline
$\rho_2 \backslash \rho_1$  & -0.6 & -0.4 & -0.2 & 0 & 0.2 & 0.4 & 0.6 \\ \hline
-0.6 & -6.619e-02 &  -5.728e-02 &  -4.887e-02 &  -3.983e-02 &  -3.114e-02 &  -2.303e-02  & -1.427e-02\\
-0.4 &  -5.191e-02 &  -4.320e-02 &  -3.409e-02 &  -2.552e-02 &  -1.726e-02 &  -9.017e-03 &  -7.615e-04\\
-0.2 &  -3.706e-02 &  -2.828e-02 &  -1.938e-02 &  -1.138e-02 &  -3.327e-03 &   4.780e-03 &   1.299e-02\\
0   &   -2.246e-02 &  -1.338e-02 &  -5.165e-03 &   2.822e-03 &   1.095e-02 &   1.877e-02 &   2.686e-02\\
0.2 &  -7.224e-03 &   1.505e-03 &   9.585e-03 &   1.776e-02 &   2.559e-02 &   3.352e-02 &   4.164e-02\\
0.4 &   8.800e-03 &   1.771e-02 &   2.568e-02 &   3.327e-02 &   4.091e-02 &   4.864e-02 &   5.639e-02\\
0.6 &   2.543e-02 &   3.414e-02 &   4.206e-02 &   4.961e-02 &   5.704e-02 &   6.453e-02 &   7.191e-02\\ \hline
\end{tabular}
\caption{Approximation errors, Set 1 for $\tau_1$, Set 2 for $\tau_2$, $T=2$. The average length of the $95\%$ confidence interval for the MC estimates is $0.0086$. }
\label{Tab3}
\end{table}

\begin{table}[t]
\centering
\begin{tabular}{cccc}
\hline
 $T$ & $g_0$ &$g_1$ & $g_2$\\ \hline
 $0.5$ &$11.3300$ &  $-0.0071$ & $0.0003$ \\ 
 $2$    &$22.4224$ &  $-0.0435$ & $0.0370$ \\ 
\end{tabular}
\caption{Contribution of zero-th and first order terms in the expansion approximation with Set 1 for $\tau_1$ and Set 2 for $\tau_2$. The corresponding default-free prices according to the B\&S formula are $c_{BS}(X_0,0,\bar{r}_s,\sigma)=11.2685$ ($T=0.5$) and $c_{BS}(X_0,0,\bar{r}_s,\sigma)=22.3480$ ($T=2$).}
\label{Tab4}
\end{table}

\begin{table}[t]
\centering
\begin{tabular}{c|ccc|ccc|ccc}
\hline
 & & $K=90$ & & &$K=100$ & & & $K=110$ &  \\
  & $\alpha=0$ & $\alpha=0.5$  & $\alpha=1$ & $\alpha=0$ & $\alpha=0.5$ & $\alpha=1$ & $\alpha=0$ & $\alpha=0.5$ & $\alpha=1$ \\ \hline
$g_0$ &$16.3455$ & $16.4559$ & $16.5663$ &$11.2208$ &$11.2965$ &$11.3723$ &$7.4639$ &$7.5142$  &$7.5646$ \\
$g_1$ &$-0.0317$  & $-0.0155$ & $0.0007$   &$-0.0254$  &$-0.0124$ &$0.0006$   &$-0.0193$&$-0.0094$ &$0.0004$ \\
$g_2$ &$0.0004$   & $0.0004$  & $0.0003$   &$0.0004$   &$0.0003$   &$0.0002$   &$0.0003$ &$0.0002$  &$0.0001$ \\\hline
\end{tabular}
\caption{Values of the zero-th and first order terms of the expansion approximation with different strikes $K$ and levels of collateralization $\alpha$.}
\label{Tab5}
\end{table}

\begin{table}[t]
\centering
\begin{tabular}{c|ccccc}
\hline
 $m$ & $-0.2$ & $-0.1$ & $0$ & $0.1$ & $0.2$ \\  \hline
  & & & $\alpha=0$ & & \\ \hline
 $c^{const}$ &  5.5458  &  8.3127 &   11.9943  & 16.7047  & 22.5212 \\
 $g_0$ &  5.2034    &   7.7995    & 11.2539      & 15.6736      & 21.1312    \\ \hline
 & & & $\alpha=0.5$ & & \\ \hline
 $c^{const}$ & 5.5728  &  8.3532 &  12.0527 &  16.7862 &  22.6310 \\
 $g_0$ & 5.2307     & 7.8405     &  11.3130   &  15.7561  &  21.2423  \\ \hline
   & & & $\alpha=1$ & & \\ \hline
 $c^{const}$ &  5.5998  &  8.3937 &  12.1112 &  16.8676  & 22.7408 \\
 $g_0$ & 5.2581 &  7.8815 &  11.3722  & 15.8385 &  21.3535  \\ \hline
\end{tabular}
\caption{Values of the zero-th and first order terms of the expansion approximation with different moneyness $m$ and levels of collateralization $\alpha$. The prices $c^{const}$ are obtained from (\ref{ca_const}).}
\label{Tab6}
\end{table}

\end{document}